\documentclass[final,3p,times]{elsarticle}

\usepackage{amssymb}
\usepackage{amsmath}

\usepackage{dsfont}
\usepackage{caption}
\usepackage[breaklinks,colorlinks,citecolor=citcolor,urlcolor=blue,linkcolor=lcolor]{hyperref}
\usepackage{cleveref}
\usepackage[symbol]{footmisc}

\biboptions{numbers,sort&compress}

\newcommand{\dd}{\mathrm{d}}

\usepackage{xcolor}

\journal{Journal of Subatomic Particles and Cosmology}

\begin{document}

\begin{frontmatter}

\title{Sampling non-abelian lattice gauge theories with diffusion models}

\author[tuwien]{Thomas R.~Ranner\footnote[1]{
Based on the talk presented by TR at xQCD2026, Busan, Korea, 14-17 July 2026, \href{mailto:thomas.ranner@tuwien.ac.at}{thomas.ranner@tuwien.ac.at}}}
\author[bielefeld]{Gert Aarts}
\author[swansea]{Diaa E.~Habibi}
\author[tuwien]{Andreas Ipp}
\author[tuwien]{David I.~M\"{u}ller}
\author[riken,utokyo]{Lingxiao Wang}
\author[sjtu,scnt]{Wei Wang}
\author[riken,sjtu]{Qianteng Zhu}

\affiliation[bielefeld]{organization={Faculty of Physics, Bielefeld University},
             city={Bielefeld},
             postcode={D-33615},
             country={Germany}}

\affiliation[swansea]{organization={Centre for Quantum Fields and Gravity, Department of Physics, Swansea University},
             city={Swansea},
             postcode={SA2 8PP},
             country={United Kingdom}}

\affiliation[tuwien]{organization={Institute for Theoretical Physics, TU Wien},
             addressline={Wiedner Hauptstra\ss{}e 8-10/136},
             city={Vienna},
             postcode={A-1040},
             country={Austria}}

\affiliation[riken]{organization={RIKEN Center for Interdisciplinary Theoretical and Mathematical Sciences (iTHEMS)},
             city={Wako},
             state={Saitama},
             postcode={351-0198},
             country={Japan}}

\affiliation[utokyo]{organization={Institute for Physics of Intelligence, Graduate School of Science, The University of Tokyo},
             addressline={Bunkyo-ku},
             city={Tokyo},
             postcode={113-0033},
             country={Japan}}

\affiliation[sjtu]{organization={State Key Laboratory of Dark Matter Physics, Key Laboratory for Particle Astrophysics and Cosmology (MOE), Shanghai Key Laboratory for Particle Physics and Cosmology, Shanghai Jiao Tong University},
             city={Shanghai},
             postcode={200240},
             country={China}}

\affiliation[scnt]{organization={Southern Center for Nuclear-Science Theory (SCNT), Institute of Modern Physics, Chinese Academy of Sciences},
             addressline={96 South Sihuan Rd., Huicheng District},
             city={Huizhou},
             postcode={516000},
             state={Guangdong},
             country={China}}

\begin{abstract}
We present gauge-equivariant diffusion models as samplers for lattice gauge theories (LGTs). Using lattice gauge-equivariant convolutional neural networks (L-CNNs) and the Metropolis-adjusted annealed Langevin algorithm (MAALA), we are able to accurately sample two-dimensional U(2) field configurations and show first promising results for four-dimensional SU(3). Our two-dimensional models extrapolate remarkably well to larger inverse couplings than originally trained on. We confirm this by comparing observables to analytically known values and to hybrid Monte Carlo results.
\end{abstract}


\end{frontmatter}

\section{Introduction}
Lattice quantum chromodynamics (lattice QCD) is the leading computational tool for non-perturbative investigations of the strong nuclear force \cite{FlavourLatticeAveragingGroupFLAG:2024oxs,Aarts:2023vsf}. While highly successful, state-of-the-art hybrid Monte Carlo (HMC) algorithms may face problems such as critical slowing down and topological freezing \cite{Schaefer:2010hu, Alles:1996vn,DelDebbio:2004xh}, which severely increase the computational cost of numerical simulations toward the continuum limit. 
Generative machine learning models, which have been successful in other fields such as computer vision and natural language processing, may offer new computational techniques to overcome these problems or at least improve upon traditional calculation methods. 

Some of the earliest generative models applied to lattice field theory were normalizing flows, which have been used for two-dimensional scalar fields \cite{Albergo:2019eim, DelDebbio:2021qwf}, two-dimensional and four-dimensional \cite{Kanwar:2020xzo,Boyda:2020hsi,Abbott:2023thq} gauge theory, fermionic theories \cite{Albergo:2021bna, Albergo:2022qfi, Abbott:2022zhs} and finally four-dimensional QCD \cite{Abbott:2022hkm}. The continuous formulation of normalizing flows has also been applied to scalar fields \cite{deHaan:2021erb, Gerdes:2022eve}. As a combination of normalizing flows with non-equilibrium Markov Chain Monte Carlo (MCMC) simulations, stochastic normalizing flows have been applied to lattice theories with promising early results \cite{Caselle:2022acb, Bulgarelli:2024brv, Bulgarelli:2024cqc, Bonanno:2025pdp, Bonanno:2026tle}. A complementary approach, focused on machine-learning perfect lattice actions with strongly reduced lattice artifacts, is provided by the fixed-point improvement program \cite{Holland:2024muu, Holland:2025fsa}.
More recently, diffusion models \cite{sohldickstein2015deepunsupervisedlearningusing, song2020generativemodelingestimatinggradients, ho2020denoisingdiffusionprobabilisticmodels, song2021scorebasedgenerativemodelingstochastic} have been explored as well, for scalar fields \cite{Cotler:2023lem,Wang:2023exq,Wang:2023sry, Vega:2025hgz, Kapust:2025ida, Chen:2026bqw, Tan:2026ibs, Sexty:2026aym}, fermionic systems \cite{Tomiya:2026tbc, Vega:2026lrm} and gauge theories \cite{Zhu:2025pmw, Vega:2025hgz, Vega:2025xma, Aarts:2026zzr, Alharazin:2026lcb, Komijani:2026lan, Aarts:2026yim}.

The main bottleneck in lattice calculations stems from numerically sampling independent configurations from a highly non-trivial probability distribution defined by the system's action. Traditional MCMC methods may suffer from very long autocorrelation times, particularly close to phase transitions. 
Diffusion models tackle this problem by connecting a trivial distribution that can be easily sampled with negligible autocorrelation to the non-trivial target distribution through a stochastic process. They have been shown to be closely related to stochastic quantization \cite{Wang:2023exq, Fukushima:2024oij} and the stochastic renormalization group \cite{Carosso:2019qpb, Cotler:2023lem}.

In this conference proceeding, we present our recent work on adapting diffusion models to non-abelian lattice gauge theories, as previously reported in \cite{Aarts:2026zzr}, and extended upon in \cite{Aarts:2026yim}. Specifically, we construct a gauge-equivariant score-based diffusion model for U($N_c$) and SU($N_c$) lattice gauge theories, which utilizes a lattice gauge-equivariant convolutional neural network (L-CNN) \cite{Favoni:2020reg}. We demonstrate that sampling two-dimensional U(2) and four-dimensional SU(3) lattice gauge theories is possible. Our approach allows for generalization beyond the fixed lattice sizes used during training.
Further, by applying the Metropolis-adjusted annealed Langevin algorithm (MAALA) \cite{Zhu:2025pmw}, an alternative sampling method for diffusion models, we are able to extrapolate to couplings far away from the training ensemble of our model in the two-dimensional U(2) setup. 

The code used for this work is available at \href{https://github.com/DM-QFT/DM_U2_LGT}{github.com/DM-QFT/DM\_U2\_LGT}.

\section{Diffusion models adapted to non-abelian lattice gauge theory}
Our aim is to calculate observables in non-abelian lattice gauge theory using Monte Carlo integration, 
\begin{align}
    \langle O \rangle \approx   \frac{1}{N_{\rm conf}} \sum_{i=1}^{N_{\rm conf}} O(\mathbf U_i), \qquad \mathbf{U}_i \sim p(\mathbf{U}) = \frac{e^{-S_E(\mathbf U)}}{\mathcal{Z}}, \qquad    S_E(\mathbf{U}) = \frac{\beta}{N_c} \sum_{x, \mu < \nu} \mathrm{Re} \, \mathrm{Tr}    [\mathds{1} - P_{x, \mu\nu}],
\end{align}
where the lattice field $\mathbf U$ consists of link variables $U_{x,\mu}$ which are elements of the gauge group (here U$(N_c)$ or SU$(N_c)$), with $x$ denoting position and $\mu$ direction. For sampling from the probability distribution $p(\mathbf{U})$, defined by the Wilson action $S_E(\mathbf{U})$ through the plaquette $P_{x, \mu\nu} = U_{x, \mu} U_{x+\hat\mu, \nu} U^\dagger_{x+\hat\nu, \mu}  U^\dagger_{x, \nu}$ and the inverse coupling $\beta$, we implement a diffusion model.

Diffusion models are a class of generative machine learning models which transform a simple \emph{prior} probability distribution into the desired, usually very complicated, \emph{target} distribution through a stochastic process. 
First, one defines a \emph{forward} diffusion process, which transforms the target distribution into the prior distribution. In LGT, it is described by a group-preserving Stratonovich stochastic differential equation (Stratonovich SDE) as
\begin{align} \label{eq:forward_gauge}
    \dd U^{(t)}_{x,\mu} = i \sum_a T^a \left[ K^a_{x,\mu}(\mathbf U^{(t)}) \, \dd t + g(t) \, \dd  W^{(t),a}_{x,\mu}\right] \! \circ  U^{(t)}_{x,\mu}, \qquad t\in[0, T],
\end{align}
with the generators of the group $T^a$, $\mathrm{Tr}[T^a T^b] = \delta^{ab}/2$, and the sum over the color index, $1\le a \le D_A$, where $D_A = N_c^2 - 1$ for SU$(N_c)$ and $D_A = N_c^2$ for U$(N_c)$. Here, $K^a_{x,\mu}$ is a gauge-covariant drift term, $g(t)$ is a time-dependent diffusion coefficient, and $W^a_{x,\mu}$ denote independent Wiener increments. At the initial diffusion time $t=0$, the links are drawn from the target distribution $p_0(\mathbf U)$. Solving the SDE evolves the probability distribution towards the prior $p_T(\mathbf U)$, whose details depend strongly on the drift and diffusion terms. For properly chosen diffusion parameters, the prior distribution can be very simple, e.g.,~the Haar measure.

The main insight used in diffusion models is that this probability evolution can be reversed through another diffusion process, the \emph{reverse} diffusion process, described by
\begin{align}\label{eq:reverse_gauge}
     \dd U^{(t)}_{x,\mu} =  i \sum_a T^a \big[ \big( \! \!K^a_{x,\mu}(\mathbf U^{( t)}) \! - \! g( t)^2 D^a_{x,\mu} \ln  p_{ t} (\mathbf U^{( t)}) \big) \dd t
    + g( t) \dd  \bar{W}^{( t), a}_{x,\mu} \big] \circ U^{( t)}_{x,\mu},
\end{align}
where $D^a_{x,\mu}$ is the group derivative.
Now $t$ runs from $T$ to $0$, $\dd \bar{W}$ denotes the Wiener process in reversed time, and the initial condition at $t=T$ is drawn from the prior distribution $p_T(\mathbf U)$. Besides the parameters of the forward diffusion, the reverse process also depends on the \emph{score} function, $D^a_{x,\mu} \ln p_t(\mathbf U)$, which is in general unknown, but will be approximated by a neural network. Solving the reverse SDE for a given $\mathbf U_T \sim p_T(\mathbf U)$ from $T$ to $0$ is equivalent to sampling directly from the target distribution $p_0(\mathbf U)$. 

To learn the score function with a neural network $s_{\theta;x,\mu}^a(\mathbf{U}, t)$, we use the denoising score matching \cite{6795935} loss function adapted to gauge fields
\begin{equation}
    \label{eq:gauge_loss_function}
    \mathcal{L} = \int_0^T  \!\! {\dd t} \,  \lambda(t) \!\!\!\!\!
    \underset{ \substack{ p_0(\mathbf{U}^{(0)}) \\ p_{0t}(\mathbf{U}^{(t)}|\mathbf{U}^{(0)}) }}
    {\mathbb{E}} 
       \sum_{x,\mu,a} \big[ s_{\theta;x,\mu}^a(\mathbf{U}^{(t)} \!\!, t) \! - \! D^a_{x,\mu} \! \ln p_{0t}(\mathbf U^{(t)} | \mathbf{U}^{(0)}) \big]^2,
\end{equation}
where $p_{0t}$ describes the finite-time transition probability of the forward process. The loss function contains an expectation over the target distribution, which, in practice, is approximated via Monte Carlo integration using a training ensemble.

\subsection{Practical setup}

For our experiments we set $K^a_{x,\mu}(\mathbf U^{(t)})$ to zero, and $g(t)=s^t$ with $s=25$. This choice is termed variance-expanding and leads to a prior distribution which is approximately the Haar measure at $t=T=1$. Instead of solving the SDE iteratively,  we use an approximation for the diffusion process over finite times in a single update step
\begin{align} \label{eq:single_step_diffusion_ve}
    {U}^{(t)}_{x,\mu}  \approx \exp \big[ i \sigma(t) \sum_a T^a  \eta^a_{x,\mu}  \big] {U}^{(0)}_{x,\mu}, \quad \sigma(t) = \sqrt{\frac{s^{2 t} - 1}{2 \ln s}},
\end{align}
with $\eta^a_{x,\mu}$ being independently drawn Gaussian random variables with zero mean and unit variance. We note that this update rule is only exact for abelian gauge groups, though we find it to be sufficiently accurate in practice. From this update rule follows the approximation of the loss function we use,
\begin{equation}
    \mathcal{L} \approx \int_0^T  \! {\dd t} \, \lambda(t) 
    \underset{ \substack{ p_0(\mathbf{U}^{(0)}) \\ p_{\eta}(\eta) }}
    {\mathbb{E}}
       \sum_{x,\mu,a} \big[ s_{\theta;x,\mu}^a(\mathbf{U}^{(t)} \!\!, t) + \frac{\eta_{x,\mu}^a}{\sigma(t)} \big]^2,
\end{equation}
where we set $\lambda(t)=\sqrt{\sigma(t)}$.
Given a training data set of field configurations from the target distribution $ p_0(\mathbf{U}^{(0)})$ and using the single-step update rule to obtain $\mathbf{U}^{(t)}$, this loss function can be evaluated very efficiently. The approximation to the SDE solution and the loss function can be understood as neglecting non-abelian effects.

\subsection{Metropolis-adjusted annealed Langevin algorithm}
Besides solving the reverse SDE we also consider a different sampling method, the Metropolis-adjusted annealed Langevin algorithm (MAALA) \cite{Zhu:2025pmw}. It is based on the observation that the score function $D^a_{x,\mu} \ln  p (\mathbf U)$ of a lattice field theory $p(\mathbf{U}) = \exp(-S_E(\mathbf{U}))/\mathcal{Z}$ is given by the negative group derivative of the action, $-D^a_{x,\mu} S_E(\mathbf{U})$. We now connect this to stochastic quantization \cite{Parisi:1980ys, Damgaard:1987rr}, a standard sampling algorithm for lattice field theory where one solves the following SDE (also called Langevin equation):
\begin{align} \label{eq:gauge_sq}
    \dd U^{(\tau)}_{x,\mu} = i \sum_a T^a \left[ -D^a_{x,\mu} S_E(\mathbf{U}^{(\tau)}) \, \dd \tau + \sqrt{2} \, \dd  W^{(\tau),a}_{x,\mu}\right] \! \circ  U^{(\tau)}_{x,\mu}.
\end{align}
In the stationary limit $\tau \rightarrow\infty$, field configurations will follow the probability distribution given by the action in the drift term, $p(\mathbf{U}) = \exp(-S_E(\mathbf{U}))/\mathcal{Z}$. By interpreting the learned time-dependent score function as a derivative of some time-dependent action, we can directly insert the learned score function evaluated at an arbitrary diffusion time $t$ into \cref{eq:gauge_sq}. The idea of MAALA is to define a grid of diffusion times $[0,t_1,\ldots,t_M]$, with $t_M$ close to $T$. Starting with samples drawn from the prior distribution, we numerically solve the Langevin equation
\begin{equation}
\label{eq:maala_gauge_sde}
     U^{(t)}_{(\tau+\epsilon)}=
    \exp\!\left[ i \sum_a T^a \left( \epsilon \, s^a_{\theta;x,\mu}(\mathbf{U}^{(t)}_{(\tau)}, t)
    +\sqrt{2\epsilon} \,  \eta^a_{x,\mu}(\tau) \right) \right]\,   U^{(t)}_{(\tau)},
\end{equation}
up to  some defined Langevin time $\tau = \mathcal{T}$ with the score function evaluated at diffusion time $t=t_M$. The result is used as an initial condition for another Langevin trajectory at $t_{M-1}$. This is repeated until reaching $t_0=0$, where the score function corresponds to the target probability. 
Assuming the neural network is trained properly, and the time grid is sufficiently fine, samples in this generative method follow the same probability path as when sampling the reverse SDE. This method is very similar to (non-machine-learned) annealed sampling, where the action is varied according to a predefined schedule.
As one MAALA trajectory returns one configuration, and we generate multiple configurations by running trajectories in parallel, the results exhibit no autocorrelation by construction. 
The key advantage of MAALA over solving the reverse SDE is the possibility of applying Metropolis adjustment using the target action, which renders the method (asymptotically) exact. Metropolis adjustment is applied on the last $5\%$ of diffusion time, where the learned score is expected to be close to the target distribution. 
Further, MAALA allows for \emph{score rescaling}, where one can sample at inverse couplings $\beta$ distinct from the $\beta_0$ of the training ensemble by rescaling the neural network as $\beta/\beta_0 \, s^a_{\theta; x, \mu}(\mathbf{U}, t)$.

\section{Numerical results}
\subsection{Two-dimensional U(2) theory}
\begin{figure}[tbp]
  \centering
  \begin{minipage}[t]{0.48\textwidth}
    \centering
    \includegraphics[width=\linewidth]{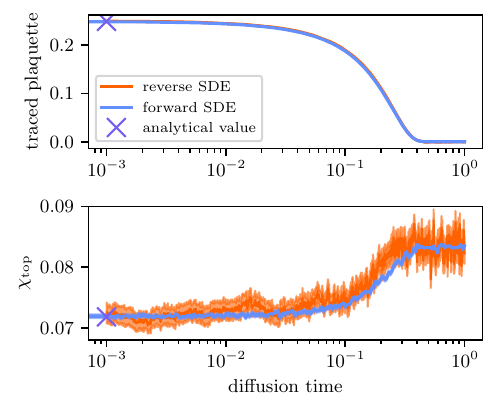}
    \captionof{figure}{The traced plaquette and the topological susceptibility calculated along the forward diffusion process and the learned reverse diffusion process for $\beta=2$, $L=16$. The trajectories show good agreement across diffusion time. At $t=1$, the distribution corresponds to the strong coupling limit with vanishing plaquette and $\chi_\mathrm{top} = 1/12$. At $t=0$, the analytical values are reproduced accurately.}
    \label{fig:reverse_sde_process}
  \end{minipage}\hfill
  \begin{minipage}[t]{0.48\textwidth}
    \centering
    \includegraphics[width=\linewidth]{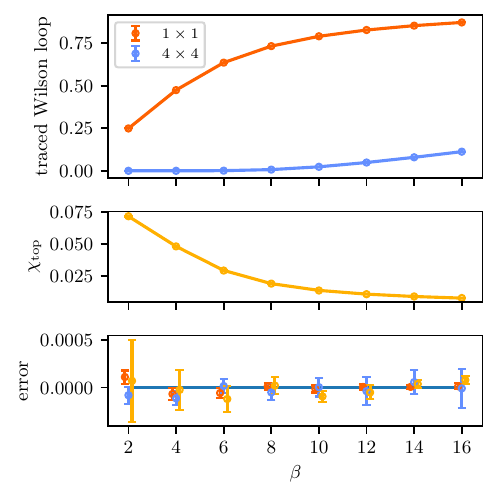}
    \captionof{figure}{Observables calculated on ensembles generated for different inverse couplings using MAALA with score rescaling. The model was trained at $\beta_0=2$. Excellent agreement between model (circles) and the analytical calculation (lines) is observed. The bottom panel shows the mismatch between predicted and analytical values expressed as absolute errors.
    }
    \label{fig:loops_and_chi_abs}
  \end{minipage}
\end{figure}

 We mainly focus on two-dimensional U(2) theory with a lattice of size $16^2$. Two-dimensional lattice gauge theory allows for exact analytical calculations of several observables, thereby enabling direct comparison to numerical results generated by the diffusion model.
We use $100 \, 000$ configurations generated with HMC as training samples at $\beta_0=2$ to train an L-CNN with two bilinear convolutional layers. In Fig.~\ref{fig:reverse_sde_process}, we compare the evolution of observables under the forward process and the learned reverse SDE. The trajectories show good agreement within statistical errors. 
Next, we investigate the extrapolation performance of MAALA sampling towards larger inverse couplings. As shown in Fig.~\ref{fig:loops_and_chi_abs} we find excellent agreement between observables of the generated ensembles and the analytical values at a fixed number of Langevin steps. We note that the number of Langevin steps necessary for a correct result increases with $\beta$, see \cite{Aarts:2026yim} for details.

\begin{figure}[tbp]
    \centering
    \includegraphics[width=0.97\linewidth]{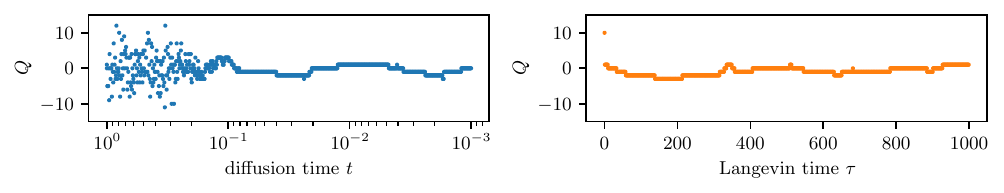}
    \caption{Evolution of the topological charge of a single configuration during the generative process for $\beta=14$, $L=16$. On the left, we use MAALA, on the right, stochastic quantization. For both trajectories, we use the same number of Langevin update steps with the same step size.}
    \label{fig:topo_freezing}
\end{figure}

Finally, in Fig.~\ref{fig:topo_freezing} we provide some intuition as to why learned annealed sampling with MAALA can be beneficial. Plotting the evolution of the topological charge during the generative process, we find that with MAALA there are frequent changes during the initial stages, while for standard stochastic quantization changes in $Q$ are suppressed by topological freezing along the whole trajectory. The rapid exploration of the phase space during initial times improves convergence towards the correct result. 

\subsection{Four-dimensional SU(3) theory}
\begin{figure}
    \centering
    \includegraphics[width=0.97\linewidth]{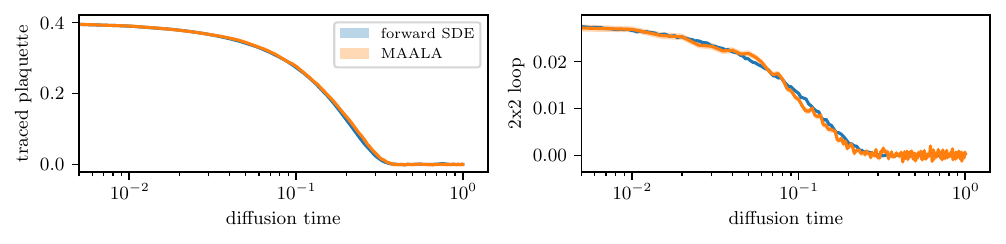}
    \caption{Comparison of the forward process with the reverse MAALA process for four-dimensional SU(3) lattice theory, $L=4$, $\beta=5$. }
    \label{fig:4d_su3_maala}
\end{figure}

As a proof of concept, we briefly investigate sampling on a four-dimensional SU(3) lattice theory with $L=4$. Here we use $5 \, 000$ configurations generated with HMC at $\beta=5$ to train an L-CNN with three bilinear convolutional layers. In Fig.~\ref{fig:4d_su3_maala} we show that during sampling with MAALA the generative process agrees with the forward diffusion process in terms of the plaquette and the $2\times 2$ loop, demonstrating that the diffusion model is able to learn sampling also in this more complicated setup.

\section{Conclusion}
We have extended the framework of diffusion models to non-abelian lattice gauge theories, showing that we can generate correct ensembles by solving the reverse SDE. As the reverse SDE is not guaranteed to be exact, we also applied the MAALA sampling method, which is (asymptotically) exact, and further allows for extrapolation to different couplings. With a fixed sufficiently high Langevin trajectory length, we find excellent extrapolation performance to a wide range of larger inverse couplings.
 While our method exhibits no autocorrelation between samples by construction, we found an effect akin to critical slowing down along the individual Langevin trajectories, leading to extended thermalization times.
 These results were achieved on a two-dimensional U(2) setup, but we also provide a first demonstration that sampling with MAALA is possible for four-dimensional SU(3) theory. Proper extrapolation results and performance comparisons for four-dimensional SU(3) theory are left to future work. More generally, it remains an open question how diffusion models perform on realistic lattice QCD setups. While the feasibility seems plausible, the performance remains unclear. 

\section*{Acknowledgments} 
We thank the DEEP-IN working group at RIKEN iTHEMS for support in the preparation of this proceeding.
GA was supported by STFC grant ST/X000648/1 and by a Royal Society Leverhulme Trust Senior Research Fellowship, and is currently supported by the W\"ubben Stiftung Wissenschaft. 
DEH is supported by the UKRI AIMLAC CDT EP/S023992/1. 
AI, DM and TR acknowledge funding from the Austrian Science Fund (FWF) 10.55776/PAT3667424.
LW is supported by JSPS KAKENHI Grant No.~25H01560, and JST-BOOST Grant No.~JPMJBY24H9. LW and QZ are also supported by the RIKEN-TRIP initiative (RIKEN-Quantum). 
WW and QZ are supported in part by the Natural Science Foundation of China under Grants No.~12125503 and No.~12305103. 
The computational results have been achieved using the Austrian Scientific Computing (ASC) infrastructure.

\setlength{\bibsep}{0.1ex}   
\def\bibfont{\footnotesize}
\bibliographystyle{elsarticle-num}
\bibliography{xqcd2026_template}

\end{document}